# Temperature drift rate for nuclear terms of NV center ground state Hamiltonian

Vladimir V. Soshenko[1], Vadim V. Vorobyov[1],
Olga Rubinas[2], Boris Kudlatsky[2,3], Anton I. Zeleneev[2,3], Stepan V. Bolshedvorskii[1,2], Vadim N. Sorokin[1,3], Andrey N. Smolyaninov[4], and Alexey V. Akimov[1,3,5]

[1]P. N. Lebedev Physical Institute, 53 Leninskij Prospekt, Moscow, 119991, Russia
[2]Moscow Institute of Physics and Technology, 9 Institutskiy per., Dolgoprudny, Moscow Region, 141700, Russia
[3]Russian Quantum Center, 100 Novaya St., Skolkovo, Moscow, 143025, Russia
[4]Photonic Nano-Meta Technologies, The Territory of Skolkovo Innovation Center, Str. Nobel b.7, Moscow, 143026, Russia
[5]Texas A&M University, 4242 TAMU, College Station, TX 77843, USA

Nitrogen-vacancy (NV) center in diamond was found to be a powerful tool for various sensing applications. The main "work horse" of this center so far has been optically detected electron resonance. Utilization of the nuclear spin has the potential of significantly improving sensitivity in rotation and magnetic field sensors. Ensemble-based sensors consume quite a bit of power, thus requiring an understanding of temperature-related shifts. In this article, we provide a detailed study of the temperature shift of the hyperfine components of the NV center.

The use of nitrogen-vacancy (NV) color center in diamond has a number of important applications in bio-imaging, sensing, and bio-sensing [1–6]. It also allows long-lasting (more than one second) room-temperature, quantum memory [7]. Nuclear spin plays an important role in many of these applications; as an example, the long-lasting quantum memory utilizes nuclear spin (since it could be robust against spin bath noise) and is very weakly modified in optical transitions [8,9]. Nuclear spin's storage properties enhance sensor sensitivity [10], and nuclear spin could be used as a direct sensing element, for example, for rotation sensing [11,12]. Its applications range from enhancing magnetic resonance imaging sensitivity in nanodiamond-based, biocompatible sensors via dynamical nuclear polarization [3,13,14] to non-volatile quantum memory [15] or room temperature quantum registers [7,8,16]. While the most impressive experimental results were achieved with $^{13}$C nuclear spin in isotopically pure diamond, nitrogen-related spin also has great potential for ensemble-based measurements since it has a definite location within the NV color center. In particular, rotation sensors benefit from this advantage [11,12].

Sensing applications require a deep understanding of the systematic shift of spin-related energy levels. This is important in both estimating systematics and realization of sensors, i.e., for spin initialization, manipulation, and readout [9,11,12,17]. Practical, ensemble-based sensors typically use very high optical pumping powers [18,19] to get high signal-to-noise ratio, leading to considerable heating of the diamond.

Thus, if not compensated, temperature shifts may cause failure of the spin manipulation and overall sensor confusion. While the diamond temperature could be monitored with the same NV center [20–22] using temperature dependence of the optically detected magnetic resonance (ODMR) [20,23,24], feedback on the nuclear spin resonance requires an understanding of behavior of the nuclear-spin-related levels with the temperature.

In this paper, we experimentally investigate a temperature shift of nitrogen ($^{14}$N) related spin in NV color center in diamond.

Measurement of nuclear spin sensitivity to temperature was done via detection of ODMR in an ensemble of NV centers (see Figure 1A). In the presence of the nitrogen nuclear spin ODMR resonance with defined component of electron spin $m_s = 0, \pm 1$ splits into 3 components (Figure 1B). The splitting is nevertheless different in the ground $m_s = 0$ and excited $m_s = -1$ states. This different splitting allows one to address nuclear sublevels independently in either the excited or ground state. Thus, if ODMR resonance is observed at microwave (MW) field frequency, corresponding to the particular hyperfine component, the strength of the resonance may be modified if transitions between hyperfine components are induced by the radio frequency (RF) field. The maximum effect of the RF field on ODMR transitions happens when the RF field is exactly at resonance with RF transitions, thus allowing measurement of the RF transition frequency by effect of RF on ODMR resonance.

To detect the position of the nuclear and electron resonances, we applied a pulsed sequence, illustrated in Figure 1C,D. The main cycle of the sequence is similar to the nuclear recursive polarization sequence [9,17]. The NV center first was prepared by a green optical pump in the ground state in a $m_s = 0$ manifold and then was transferred into a $m_s = -1$ manifold using the MW $\pi$ pulse (see Figure 1C). Then, RF pulse was applied, which, if resonant, mixed hyperfine levels of the $m_s = -1$ manifold. The RF pulse is followed by an optical pulse that returns most of the population into the $m_s = 0$ manifold, but doesn't much change the nuclear spin population. Then, another microwave $\pi$ pulse was applied. As a result of such a cycle, part of the population affected by the RF transition will no longer participate in the second MW transition. Thus, when the RF field is resonant with the optical detected magnetic resonance (ODMR), the signal will be suppressed.

To determine the center of the ODMR transition, MW frequency is swept around the transition frequency (see Figure 1D). This scan was performed for each RF frequency (see Figure 2A) thus allowing a view of both MW and RF resonance positions at the same time (see Figure 2B,C).

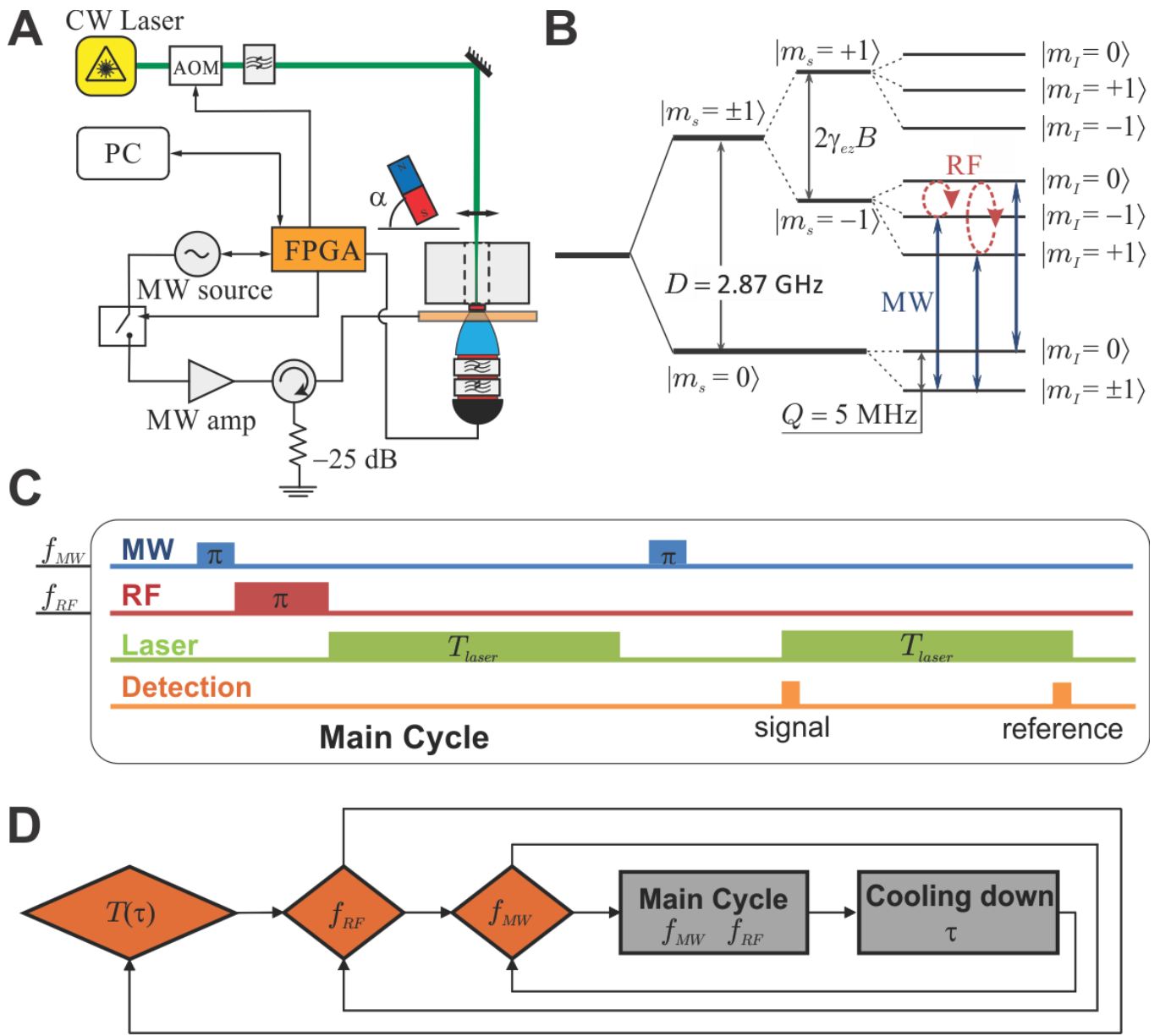

Figure 1. A) Experimental setup for nuclear spin spectroscopy on an NV centers ensemble. An FPGA stay for a Field Programmable Gate Array was used to control pulse sequence AOM – acousto optic modulator. B) Energy levels of the ground state of an NV center. $D$ stays for zero field splitting, $Q$ – for hyperfine structure constant for splitting due to $^{14}$N nuclear spin, $2\gamma_{es}B$ – Zeeman splitting between electron spins. The blue solid arrows show the allowed dipole MW transitions to the $m_s = -1$ manifold, and the red circular dashed lines are the once–allowed RF transitions of nuclear spin in the $m_s = -1$ manifold. C) Main cycle of the pulsed scheme. D) Experimental sequence.

The ODMR resonance position is known to have a ~70 kHz/K spectral shift, which allows one to measure the temperature of the sample [20,21]. This enables us to calibrate the temperature of the sample for different cooling-down cycle lengths (Figure 2D). The idea here is explained as follows. At large optical powers, absorption of the diamond plate due to the presence of color centers could lead to quite significant heating of the plate, determined by both the heat-sink cooling rate of the plate and the optical power that is dissipated into heat within it. In our case, the heating was up to about 60 degrees with continuous-wave

(CW) laser excitation. Thus, to vary the temperature of the sample, it was enough to vary the duty cycle of the laser in the sequence that was applied.

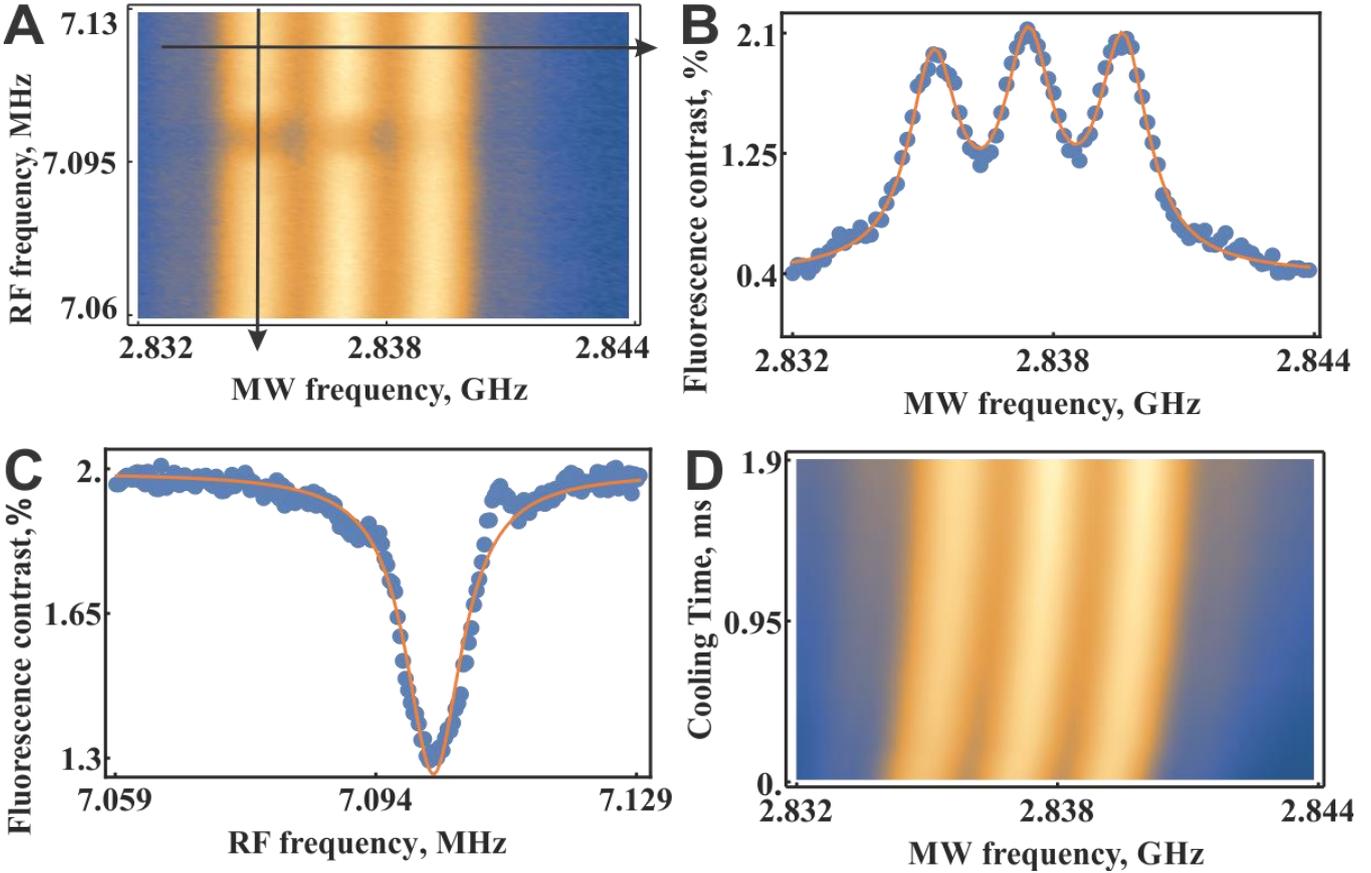

Figure 2. A) 2D scan of the RF and MW resonances. The plot averaged 6 cycles with the same cooling-down time. B) Cross-section of the 2D plot at A with $f_{RF} = 7.128$ MHz, corresponding to the MW resonances not affected by RF. The solid line corresponds to the fit with 3 Lorenzian curves. C) Cross-section of 2D plot at A along the maximum of the left peak at B. The solid line corresponds to the Lorenzian fit. D). The ODMR of NV ensemble versus cooling-down time. The position of resonance slightly blue shifted with an increase of the cooling-down time, which correspond to the decrease in the sample temperature.

Calibration of the temperature versus cooling time is demonstrated in Figure 3A. Here, all three ODMR resonances (see Figure 2B) were fitted with the three-Lorentzian fit and the position of the central peak was chosen for Figure 3A. Its shift was converted to a temperature-relative shift using the known value of the NV ODMR temperature shift [20] for each cooling time. Though dependence is not linear, a one-to-one correspondence can easily be established.

Once calibration was achieved, the shift of nuclear transitions with temperature was easy to measure (see Figure 3B). Here, to calculate the position of the RF resonance, the spectrum of the RF resonance was fitted with a single Lorenzian curve and the center of this curve was taken as the resonance position (see Figure 2C). Clearly, the shift is linear with the temperature and toward the smaller frequencies. While the magnitude of the shift is not large, it is still quite significant in comparison to the resonance width, which in our case was about 8 kHz.

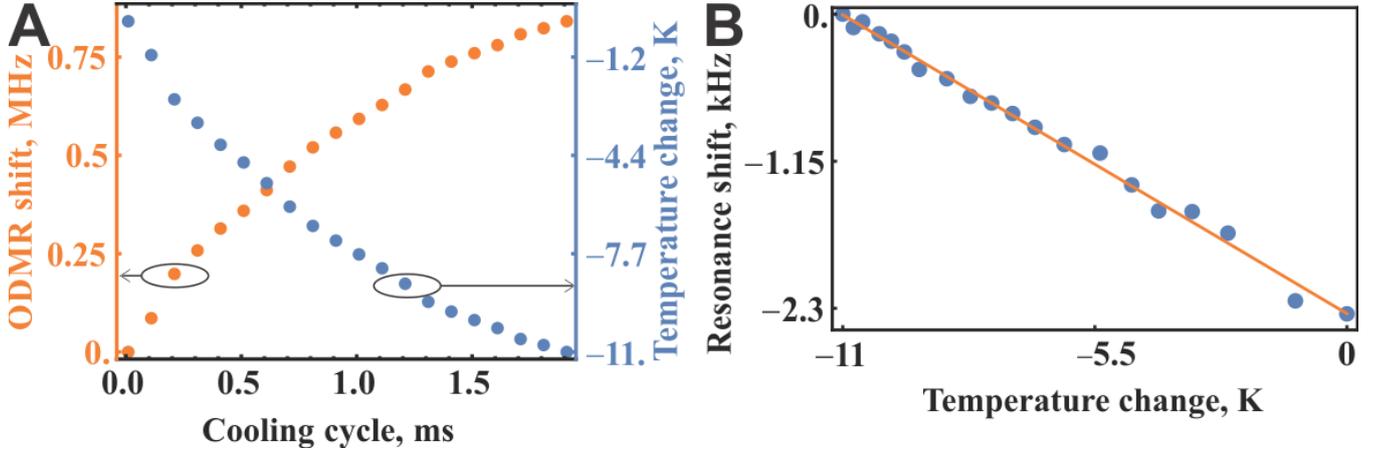

Figure 3. A) The temperature of the sample recalculated from shift of the NV ensemble ODMR position. The shift is calculated with respect to the 2.837 GHz MW frequency. B) Dependence of nuclear resonance transition frequency detuning on temperature change of the diamond. Clearly, a linear dependence is observed. The detuning is calculated with respect to the 7.1028 MHz RF frequency.

The NV center excited state has two allowed NMR transitions (see Figure 4A). The shifts of the resonance frequency of these two transitions are opposite, as seen in Figure 4B. Still, the magnitudes of the slopes of these two transitions are not exactly the same, thus making it possible to separate the contributions of the shift $Q$ due nuclear spin electric quadrupole moment and the hyperfine structure constant $A_\parallel$ into the levels shift. The interaction Hamiltonian $H$ for the nuclear spin in NV center is [25]:

$$H = QI_z^2 + S_z A_\parallel I_z + \gamma_n B_z I_z \qquad (1)$$

Here, $B_z$ is external magnetic field, $\gamma_n$ – nuclear spin gyromagnetic ratio, $I_z$ – nuclear spin, $S_z$ – electron spin. Since the electronic spin in the ground state is 0 (see Figure 4A), the nuclear state splitting in the ground state is independent from $A_\parallel$. To the contrary, in the excited state both $A_\parallel$ and $Q$ contribute to the transition. The hyperfine structure constant contribution to the transition shift must be symmetric with respect to $m_I = \pm 1$, and, therefore, the asymmetry of the excited state shift may only be due to the quadrupole term. The total $\delta f_{RF}$ shift for the single RF transition is thus:

$$\delta f_{RF}(T) = \frac{1}{h}\left(\frac{\partial Q}{\partial T}\delta T \pm \frac{\partial A_\parallel}{\partial T}\delta T\right) \qquad (2)$$

Where "+" is for NMR2, and "−" for NMR1 transitions respectively. And from measured values $\partial f_{NMR1}/\partial T = 163 \pm 5$ Hz/°C and $\partial f_{NMR2}/\partial T = -213 \pm 4$ Hz/°C one could extract:

$$\frac{1}{h}\frac{\partial A_\parallel}{\partial T} = \frac{1}{2}\left(\frac{\partial f_{NMR2}}{\partial T} - \frac{\partial f_{NMR1}}{\partial T}\right) = -188 \pm 4 \frac{\text{Hz}}{\text{°C}} \quad (3)$$

and

$$\frac{1}{h}\frac{\partial Q}{\partial T} = \frac{1}{2}\left(\frac{\partial f_{NMR2}}{\partial T} + \frac{\partial f_{NMR1}}{\partial T}\right) = -24 \pm 4 \frac{\text{Hz}}{\text{°C}} \quad (4)$$

Nevertheless, the variation of temperature sensitivity is different for diamond around 10%, which is the final uncertainty of our measurements. The dominating role of the hyperfine term is the temperature sensitivity and most likely is due to the slight deformation of the electron orbitals with temperature. The hyperfine term $A_\parallel$ could be expressed as the following [25]:

$$A_\parallel = f_A + 2a_{A_1},$$

Where $f_A$ is the Fermi contact term and $a_A$ is a dipole-dipole nuclear-electron interaction, which was estimated as $-h \cdot 2.5$ MHz and $h \cdot 187$ kHz respectively [25,26]. Thus $f_A$ is the dominant term in the hyperfine coupling. The expression for $f_A$ is:

$$f_A = \gamma_{el}\gamma_n \frac{\mu_0}{4\pi\hbar^2}\left|\psi^{el}_{@n}\right|^2,$$

Where $\gamma_{el}, \gamma_n$ are the nuclear and electron gyromagnetic ratio, $\left|\psi^{el}_{@n}\right| = \langle e|\delta(r - r_N)|e\rangle$ is an electron spin density at $^{14}$N nuclear site. We note that it was speculated that dependence of spin density from distance is exponential [27,28] Hence, dependence of $A_\parallel$ on temperature is the following:

$$\frac{dA_\parallel}{dT} = \frac{dA_\parallel}{dr}\frac{dR}{dT} = \frac{A_\parallel}{r_0}\frac{dR}{dT}$$

Where $R$ is the distance between nitrogen and first near carbon atom and equal to 0.252 nm, $r_0$ is the decay constant of the electron spin density, and $dR/dT$ is equal to $-2.52(2) \times 10^{-5}$ K$^{-1}$ [20]. Taking $r_0$ as 0.3 nm (characteristic length of the diamond lattice) we obtained that $\frac{dA_\parallel}{dT} \simeq -211.12$ Hz/K, which is within 10% with our experimental result. The remaining discrepancy could be due to the dipole-dipole interaction and thus needs further investigation.

We note that the quadrupole shift is an order of magnitude lower than the hyperfine interaction term. The quadrupole shift is likely due to the change of the electric field gradient seen by the nuclear spin with expansion of the sample [29]. However, to our knowledge, the fraction of $p$ orbitals into the nuclear spin site is low [29], which might be the reason for the low sensitivity to temperature.

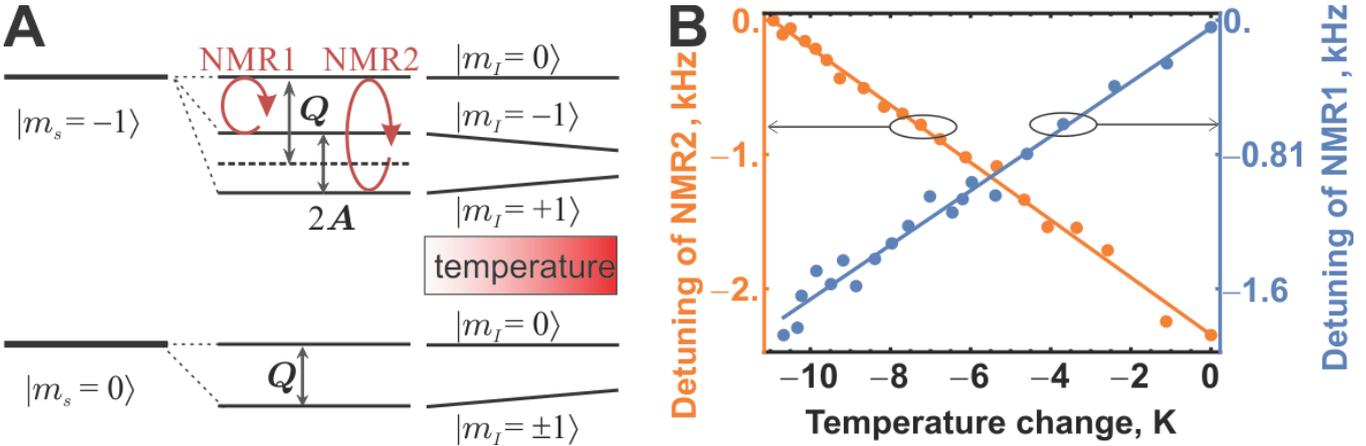

Figure 4. A) Schematic representation of the level structure shift due to temperature. NMR1 and NMR2 transitions are notated with circular arrows. B) Shift of NV hyperfine transition lines NMR1 and NMR2 due to the change in temperature (experiment). The detunings are calculated from 7.1028 MHz in case of NMR2 and from 2.7871 MHz in case of NMR1.

The temperature-dependent shift of the $^{14}$N nuclear spin associated with the hyperfine splitting of the NV center was experimentally measured. The shift of about 200 Hz/°C turn out to be quite noticeable and must be taken in to account in sensors utilizing nuclear spin, such as gyroscopes or magnetic field sensors. The major part of the shift comes from the hyperfine constant contribution $\partial A/\partial T = -188 \pm 20$ Hz/°C and is well described by the Fermi contact term drift, while the quadrupole moment contribution $\partial Q/\partial T = -24 \pm 4$ Hz/°C is an order of magnitude less sensitive to temperature.

## ACKNOLEDGEMENT

The work was supported by Russian Science Foundation Grant #16-19-10367 (experimental measurement of the temperature shift) and the Ministry of Education and Science of the Russian Federation in the framework of increase Competitiveness Program of NUST ''MISIS", implemented by a governmental decree dated 16th of March 2013 (sample preparation).

## APPENDIX

Our experimental setup allows for a controlling spin ensemble in bulk diamond at room temperature (see Figure 1A). For a light source, we used 3 W of 532 nm laser radiation (Laser Quantum Finesse Pure 10 W) focused onto a 0.05 mm spot. The intensity of the laser beam was modulated with an acousto-optical modulator (AOM) able to hold a large optical power (Gooch-Housego, I-M110-2C10B6-3-GH26). As the sample for our research, we used a diamond plate polished perpendicular to 111 crystallographic axis (Velman LLC) with approximately 1 ppm of NV centers [30]. The NV centers' fluorescence was collected with a parabolic concentrator (Edmund optics), similar to [2,31,32]. A constant magnetic field was formed by a permanent magnet and applied along the 111 crystallographic axis.

The microwave (MW) field was formed by an antenna composed of two, 5-mm-diameter coaxial loops made out of 1 mm copper wire and separated by 6 mm with diamond sample in between [33]. These loops were terminated with parallel copper plates that formed a capacitor. This assembly forms an LC resonator, the resonance frequency of which could be tuned to the desired frequency of the NV center ODMR transition. In our experiments, this was excited with a weekly capacitive-coupled feed line. A Rohde-Schwarz SMA100A signal generator was the source of the MW signal and was modulated by switch (Minicircuits ZASWA-2-50-DR+) and then amplified using a Minicircuits ZHL-16W-43X+ amplifier. The antenna was fed through the circulator to avoid back reflections. A Stanford Research signal generator SG384 was the source of radio frequency (RF) signal. The signal was amplified with a VectaWave VBA100-30 amplifier. The amplifier was directly connected to the antenna, which was formed by 2 x 10 loops of 1 mm copper wire and terminated with a 50-ohm power terminator. Typical Rabi frequencies achieved in such a system for electron and nuclear spin were 2 MHz and 10 kHz, respectively.